\documentclass[aps,prl,showpacs,showkeys,amsmath,amssymb,
twocolumn,
floatfix,
superscriptaddress
]{revtex4-1}
\usepackage{graphicx}
\usepackage{times}
\usepackage{verbatim}
\usepackage{multirow}
\usepackage[usenames,dvipsnames]{color}

\usepackage[normalem]{ulem}
\usepackage{accents}

\newcommand{\nuebar}{$\overline{\nu}_{e}$}

\newcommand*{\pbar}[1]{\accentset{(-)}{#1}}

\begin{document}

\title{Search for a Light Sterile Neutrino at Daya Bay }

\newcommand{\ECUST}{\affiliation{Institute of Modern Physics, East China University of Science and Technology, Shanghai}}
\newcommand{\IHEP}{\affiliation{Institute~of~High~Energy~Physics, Beijing}}
\newcommand{\Wisconsin}{\affiliation{University~of~Wisconsin, Madison, Wisconsin, USA}}
\newcommand{\Yale}{\affiliation{Department~of~Physics, Yale~University, New~Haven, Connecticut, USA}}
\newcommand{\BNL}{\affiliation{Brookhaven~National~Laboratory, Upton, New York, USA}}
\newcommand{\NTU}{\affiliation{Department of Physics, National~Taiwan~University, Taipei}}
\newcommand{\NUU}{\affiliation{National~United~University, Miao-Li}}
\newcommand{\Dubna}{\affiliation{Joint~Institute~for~Nuclear~Research, Dubna, Moscow~Region}}
\newcommand{\CalTech}{\affiliation{California~Institute~of~Technology, Pasadena, California, USA}}
\newcommand{\CUHK}{\affiliation{Chinese~University~of~Hong~Kong, Hong~Kong}}
\newcommand{\NCTU}{\affiliation{Institute~of~Physics, National~Chiao-Tung~University, Hsinchu}}
\newcommand{\NJU}{\affiliation{Nanjing~University, Nanjing}}
\newcommand{\TsingHua}{\affiliation{Department~of~Engineering~Physics, Tsinghua~University, Beijing}}
\newcommand{\SZU}{\affiliation{Shenzhen~University, Shenzhen}}
\newcommand{\NCEPU}{\affiliation{North~China~Electric~Power~University, Beijing}}
\newcommand{\Siena}{\affiliation{Siena~College, Loudonville, New York, USA}}
\newcommand{\IIT}{\affiliation{Department of Physics, Illinois~Institute~of~Technology, Chicago, Illinois, USA}}
\newcommand{\LBNL}{\affiliation{Lawrence~Berkeley~National~Laboratory, Berkeley, California, USA}}
\newcommand{\UIUC}{\affiliation{Department of Physics, University~of~Illinois~at~Urbana-Champaign, Urbana, Illinois, USA}}
\newcommand{\CDUT}{\affiliation{Chengdu~University~of~Technology, Chengdu}}
\newcommand{\RPI}{\affiliation{Department~of~Physics, Applied~Physics, and~Astronomy, Rensselaer~Polytechnic~Institute, Troy, New~York, USA}}
\newcommand{\SJTU}{\affiliation{Shanghai~Jiao~Tong~University, Shanghai}}
\newcommand{\BNU}{\affiliation{Beijing~Normal~University, Beijing}}
\newcommand{\WM}{\affiliation{College~of~William~and~Mary, Williamsburg, Virginia, USA}}
\newcommand{\Princeton}{\affiliation{Joseph Henry Laboratories, Princeton~University, Princeton, New~Jersey, USA}}
\newcommand{\VirginiaTech}{\affiliation{Center for Neutrino Physics, Virginia~Tech, Blacksburg, Virginia, USA}}
\newcommand{\CIAE}{\affiliation{China~Institute~of~Atomic~Energy, Beijing}}
\newcommand{\SDU}{\affiliation{Shandong~University, Jinan}}
\newcommand{\NanKai}{\affiliation{School of Physics, Nankai~University, Tianjin}}
\newcommand{\UC}{\affiliation{Department of Physics, University~of~Cincinnati, Cincinnati, Ohio, USA}}
\newcommand{\DGUT}{\affiliation{Dongguan~University~of~Technology, Dongguan}}
\newcommand{\XJTU}{\affiliation{Xi'an Jiaotong University, Xi'an}}
\newcommand{\UCB}{\affiliation{Department of Physics, University~of~California, Berkeley, California, USA}}
\newcommand{\HKU}{\affiliation{Department of Physics, The~University~of~Hong~Kong, Pokfulam, Hong~Kong}}
\newcommand{\UH}{\affiliation{Department of Physics, University~of~Houston, Houston, Texas, USA}}
\newcommand{\Charles}{\affiliation{Charles~University, Faculty~of~Mathematics~and~Physics, Prague}}
\newcommand{\USTC}{\affiliation{University~of~Science~and~Technology~of~China, Hefei}}
\newcommand{\TempleUniversity}{\affiliation{Department~of~Physics, College~of~Science~and~Technology, Temple~University, Philadelphia, Pennsylvania, USA}}
\newcommand{\CUC}{\affiliation{Institute of Physics, Pontifical Catholic University of Chile, Santiago, Chile}} 
\newcommand{\CGNPG}{\affiliation{China General Nuclear Power Group}}
\newcommand{\NUDT}{\affiliation{College of Electronic Science and Engineering, National University of Defense Technology, Changsha}} 
\newcommand{\IowaState}{\affiliation{Iowa~State~University, Ames, Iowa, USA}}
\newcommand{\ZSU}{\affiliation{Sun Yat-Sen (Zhongshan) University, Guangzhou}}
\author{F.~P.~An}\ECUST
\author{A.~B.~Balantekin}\Wisconsin
\author{H.~R.~Band}\Wisconsin
\author{W.~Beriguete}\BNL
\author{M.~Bishai}\BNL
\author{S.~Blyth}\NTU
\author{I.~Butorov}\Dubna
\author{G.~F.~Cao}\IHEP
\author{J.~Cao}\IHEP
\author{Y.L.~Chan}\CUHK
% ...

\collaboration{The Daya Bay Collaboration}\noaffiliation
\date{\today}

\begin{abstract}
\noindent {A search for light sterile neutrino mixing was performed with the first 217 days of data from the 
Daya Bay Reactor Antineutrino Experiment. The experiment's unique configuration of multiple 
baselines from six 2.9~GW$_{\rm th}$ nuclear reactors to six antineutrino detectors
deployed in two near (effective baselines 512~m and 561~m) and one far (1579~m) underground 
experimental halls makes it possible to test for oscillations to a fourth (sterile) neutrino in the 
$10^{-3}~{\rm eV}^{2} < |\Delta m_{41}^{2}| < 0.3~{\rm eV}^{2}$ range. 
The relative spectral distortion due to electron 
antineutrino disappearance was found to be consistent with 
that of the three-flavor oscillation model. The derived limits on $\sin^22\theta_{14}$ cover the $10^{-3}~{\rm eV}^{2} \lesssim |\Delta m^{2}_{41}| \lesssim 0.1~{\rm eV}^{2}$ region, which was largely unexplored.
}
\end{abstract}

\maketitle

\par

Measurements in the past decades have revealed large mixing between the flavor and mass eigenstates of neutrinos. The neutrino mixing 
framework~\cite{Pontecorvo:1957cp, Pontecorvo:1967fh, Maki:1962mu} with three flavors has been successful in explaining most experimental results, 
and several-percent precision has been attained in the
 determination of the neutrino mixing angles and the mass splittings.
Despite this great progress, there is still room for other generations of neutrinos to exist. 
Fits to precision electroweak measurements~\cite{Beringer:1900zz, ALEPH:2005ab} have limited the number of light active neutrino flavors to three, although other light neutrinos may exist as long as they do not participate in standard V-A interactions.
These neutrinos, which arise in extensions of the Standard 
Model that incorporate neutrino masses, are typically referred to as sterile neutrinos~\cite{Pontecorvo:1967fh}.

\par

In addition to being well-motivated from the theoretical standpoint, 
sterile neutrinos are among the leading candidates to resolve 
outstanding puzzles in astronomy and cosmology. 
Sterile neutrinos with $\sim$keV masses are good candidates 
for non-baryonic Dark Matter~\cite{Dodelson:1993je,Kusenko:2009up}. 
Light sterile neutrinos with eV or sub-eV mass have been shown to help reconcile the tensions in 
the cosmological data between current measurements of the present and early Universe~\cite{Wyman:2013lza} as well as between CMB and lensing 
measurements~\cite{Battye:2013xqa}. 
The recent B-mode polarization data from BICEP2~\cite{Ade:2014xna} has spurred even more discussion in this area~\cite{Giusarma:2014zza,Zhang:2014dxk,Archidiacono:2014apa,Leistedt:2014sia}.

\par

If light sterile neutrinos mix with the three active neutrinos, their presence could be detected via the modification to the latter's oscillatory behavior. Various searches for active-sterile neutrino mixing in the mass-squared splitting $|\Delta m^2| > 0.1~{\rm eV}^2$ region have been carried out in this way. 
The LSND~\cite{Aguilar:2001ty} and MiniBooNE~\cite{Aguilar-Arevalo:2013pmq,AguilarArevalo:2007it} experiments observed excesses of electron (anti-)neutrino events in the muon (anti-)neutrino beams, which could be interpreted as sterile neutrino oscillation with $|\Delta m^2| \sim 1~{\rm eV}^2$. However, these results are in tension~\cite{Maltoni:2007zf,Karagiorgi:2009nb,Karagiorgi:2011ut,Giunti:2011cp} with the limits derived from other appearance~\cite{Armbruster:2002mp, Astier:2003gs, Agafonova:2013xsk, Antonello:2013gut} or disappearance searches~\cite{Stockdale:1984cg, Dydak:1983zq, Cheng:2012yy,Mahn:2011ea, Fukuda:2000np, Adamson:2011ku, Declais:1994su, Conrad:2011ce,Palazzo:2013bsa,Esmaili:2013yea,Girardi:2014wea}.
Moreover, a reanalysis of the measured {\it vs.}\ predicted electron antineutrino events from previous reactor experiments has revealed a deficit of about 6\%~\cite{anom,anom1}. Although the significance of this effect is still under discussion~\cite{Mueller:2011nm, Hayes:2013wra}, it is compatible with the so-called Gallium Anomaly~\cite{Hampel:1997fc,Abdurashitov:2009tn,Giunti:2010zu} in that both can be explained by introducing a sterile neutrino with $|\Delta m^2| > 0.5~{\rm eV}^2$~\cite{Giunti:2012tn}.
Until now however, the  $|\Delta m^2| < 0.1~{\rm eV}^2$ region has remained largely unexplored.

\par

This Letter describes a search for a light sterile neutrino via its mixing 
with the active neutrinos using more than 300,000 reactor antineutrino 
interactions collected in the Daya Bay Reactor Antineutrino Experiment. 
This data set was recorded during the six-detector data period
from December 2011 to July 2012.
Since the antineutrino detectors are located at baselines ranging from a few hundred to almost two thousand meters away from the reactor cores, 
Daya Bay is most sensitive to active-sterile neutrino mixing in the 
$10^{-3}~{\rm eV}^2 < |\Delta m^2| < 0.3~{\rm eV}^2$ range.
In this region, a positive signal for active-sterile neutrino mixing would predominantly manifest itself as an additional spectral distortion with a frequency different from the one due to the atmospheric mass splitting. 

\par
This work used a minimal extension of the Standard Model: the 3 (active) + 1 (sterile) neutrino mixing model. 
In this model, if the neutrino mass is much smaller than its momentum,
the probability that an \nuebar\ produced with energy $E$ is detected as
an \nuebar\ after traveling a distance $L$ is given by
\begin{equation}
P_{\overline{\nu}_e\rightarrow\overline{\nu}_e} = 1 - 4
\sum^{3}_{i=1}\sum^{4}_{j > i}|U_{ei}|^{2}|U_{ej}|^{2}\sin^{2}\Delta_{ji}.
\label{eq:psurv-4nu}
\end{equation}
\noindent Here $U_{ei}$ is the element of the neutrino mixing matrix for the flavor eigenstate $\nu_e$ and the mass eigenstate $\nu_i$, 
$\Delta_{ji}=1.267 {\Delta}m^2_{ji}({\rm eV}^2)\frac{L({\rm m})}{E({\rm MeV})}$ 
with ${\Delta}m^2_{ji} = m_{j}^2 - m_{i}^{2}$ being 
the mass-squared difference between the mass eigenstates $\nu_j$ and $\nu_i$.
Using the parameterization of Ref.~\cite{Palazzo:2013bsa}, 
$U_{ei}$ can be expressed in terms of the neutrino mixing angles 
$\theta_{14}$, $\theta_{13}$ and $\theta_{12}$:
\begin{eqnarray}\label{eq:uarray}
U_{e1} &=& \cos\theta_{14} \cos\theta_{13} \cos\theta_{12}, \nonumber \\
U_{e2} &=& \cos\theta_{14} \cos\theta_{13} \sin\theta_{12}, \nonumber \\
U_{e3} &=& \cos\theta_{14} \sin\theta_{13}, \nonumber \\
U_{e4} &=& \sin\theta_{14}.
\end{eqnarray}
\noindent If $\theta_{14} = 0$, the probability returns to the expression 
for three-neutrino oscillation. 

\par

The Daya Bay experiment has two near underground experimental halls (EH1 and EH2) and one far hall (EH3).
Each hall houses functionally identical, three-zone antineutrino detectors (ADs) submerged in pools of ultra-pure water segmented into two optically decoupled regions. The water pools are instrumented with photomultiplier tubes (PMTs) to tag cosmic-ray-induced interactions. 
Reactor antineutrinos were detected via the inverse $\beta$-decay (IBD) reaction ($\overline{\nu}_{e}+p \rightarrow e^{+}+n$).
The coincidence of the prompt ($e^+$ ionization and annihilation) and delayed ($n$ capture on Gd) signals 
efficiently suppressed the backgrounds,
which amounted to less than 2\% (5\%) of the entire candidate samples in the near (far) halls~\cite{An:2013zwz}. 
The prompt signal measured the $\overline{\nu}_e$ energy with an energy 
resolution $\sigma_E/E\approx8\%$ at 1 MeV.
More details on the reconstruction and detector performance can be found in Ref.~\cite{DayaBay:2012aa}. A summary of the IBD candidates used in this analysis, together with the baselines of the three experimental halls to each pair of reactors, is shown in Table~\ref{tab:baseline}. 

\begin{table}[htb]
\begin{center}
\caption{Total number of IBD candidates and baselines of the three experimental halls to the reactor pairs.
\label{tab:baseline}}
\begin{tabular}[c]{r|r|r|r|r} \hline\hline
\multirow{2}*{Location} & \multirow{2}*{IBD candidates} & 
\multicolumn{3}{c}{Mean Distance to Reactor Core (m)}\\\cline{3-5}
 & & \mbox{ } Daya Bay & \mbox{ } Ling Ao & \mbox{ } Ling Ao-II \\\hline
\mbox{ } EH1 & \mbox{  }203809 & 365 & 860 & 1310  \\ 
\mbox{ } EH2 & \mbox{  }92912 & 1345 & 479 & 528  \\ 
\mbox{ } EH3 & \mbox{  }41589 & 1908 & 1536 & 1541  \\ \hline \hline
\end{tabular}
\end{center}
\end{table}

\par
The uncertainty in the absolute energy scale of positrons was estimated to be about 1.5\% through a combination of the uncertainties of calibration data and various energy models~\cite{An:2013zwz}. 
This quantity had a negligible effect 
on the sensitivity of the sterile neutrino search 
due to the relative nature of the measurement with 
functionally identical detectors. 
The uncertainty of the relative energy scale was determined from the 
relative response of all ADs to various calibration sources 
that spanned the IBD positron energy range, and was found to be 0.35\%. 
The predicted $\overline{\nu}_e$ flux took into account the daily livetime-corrected 
thermal power, the fission fractions of each isotope as provided by the reactor company, 
the fission energies, and the number of 
antineutrinos produced per fission per isotope~\cite{An:2013uza}. 

The precision of the measured baselines 
was about 2 cm with both 
the GPS and Total Station~\cite{An:2012eh}.
The geometric effect due to the finite size of the reactor cores 
and the antineutrino detectors, 
whose dimensions are comparable to the oscillation length 
at $|\Delta m^{2}| \sim ~{\rm eV}^2$, 
was assessed by assuming that antineutrinos were produced and interacted uniformly in these volumes. 
The impact was found to be unimportant in the range of $\Delta m^2$ where Daya Bay is most sensitive ($|\Delta m^2| <$ $0.3$ eV$^2$). 
Higher order effects, such as the non-uniform production of 
antineutrinos inside the reactor cores 
due to a particular reactor fuel burning history, also had a negligible impact on the final result. 

\begin{figure}[!htb]
\centering
\includegraphics[width=\columnwidth]{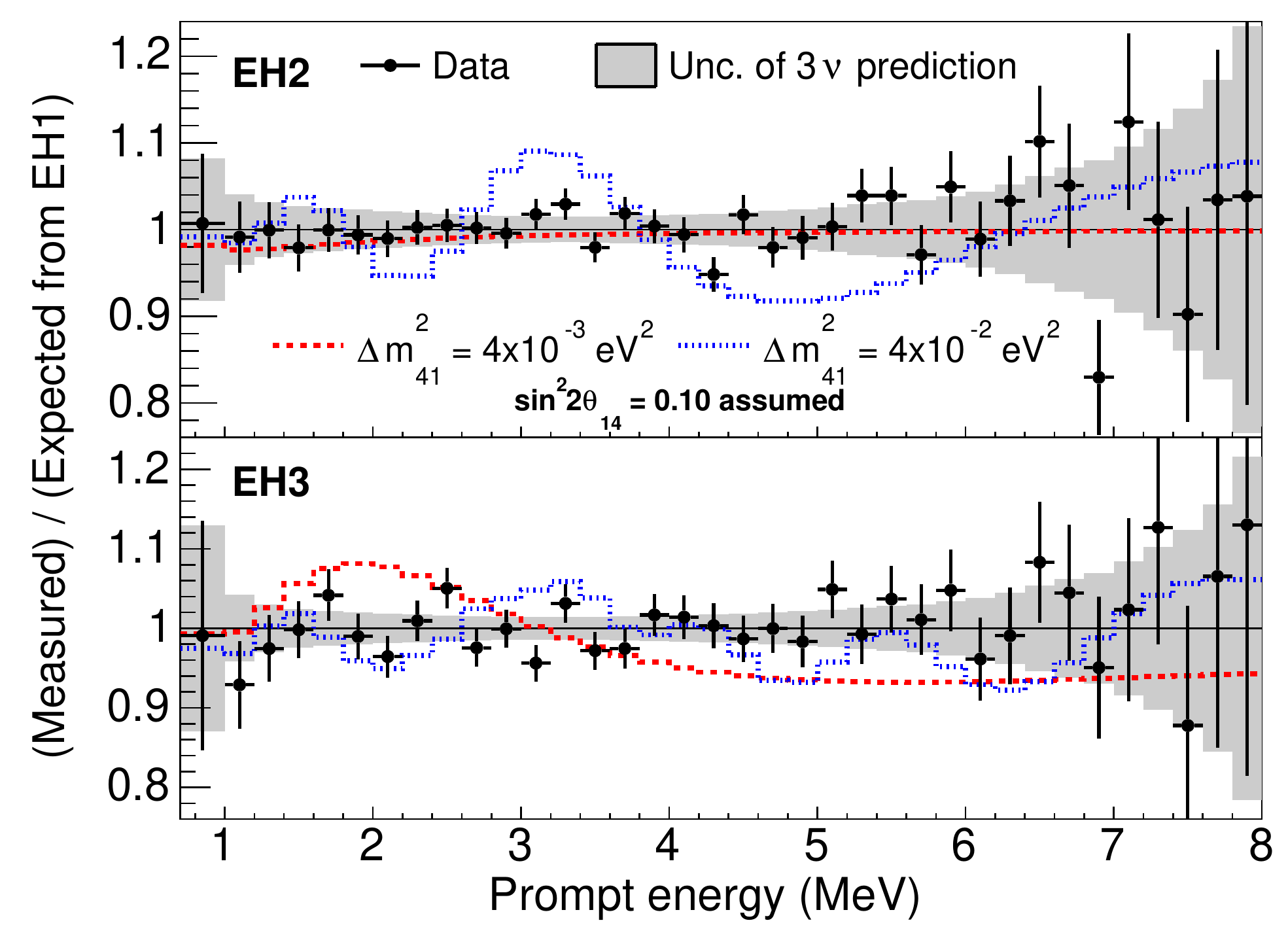}
\caption{(color online) Prompt energy spectra observed at EH2 (top) and EH3 (bottom), divided by the prediction from the EH1 spectrum with the three-neutrino best fit oscillation parameters from the previous Daya Bay analysis~\cite{An:2013zwz}. The gray band represents the uncertainty of three-neutrino oscillation prediction, which includes the statistical uncertainty of the EH1 data and all the systematic uncertainties. Predictions with $\sin^{2}2\theta_{14} = 0.1$ and two representative $|\Delta m^2_{41}|$ values are also shown as the dotted and dashed curves.}
\label{fig:SpectralRatio}
\end{figure}

\begin{figure}[!htb]
\centering
\includegraphics[width=\columnwidth]{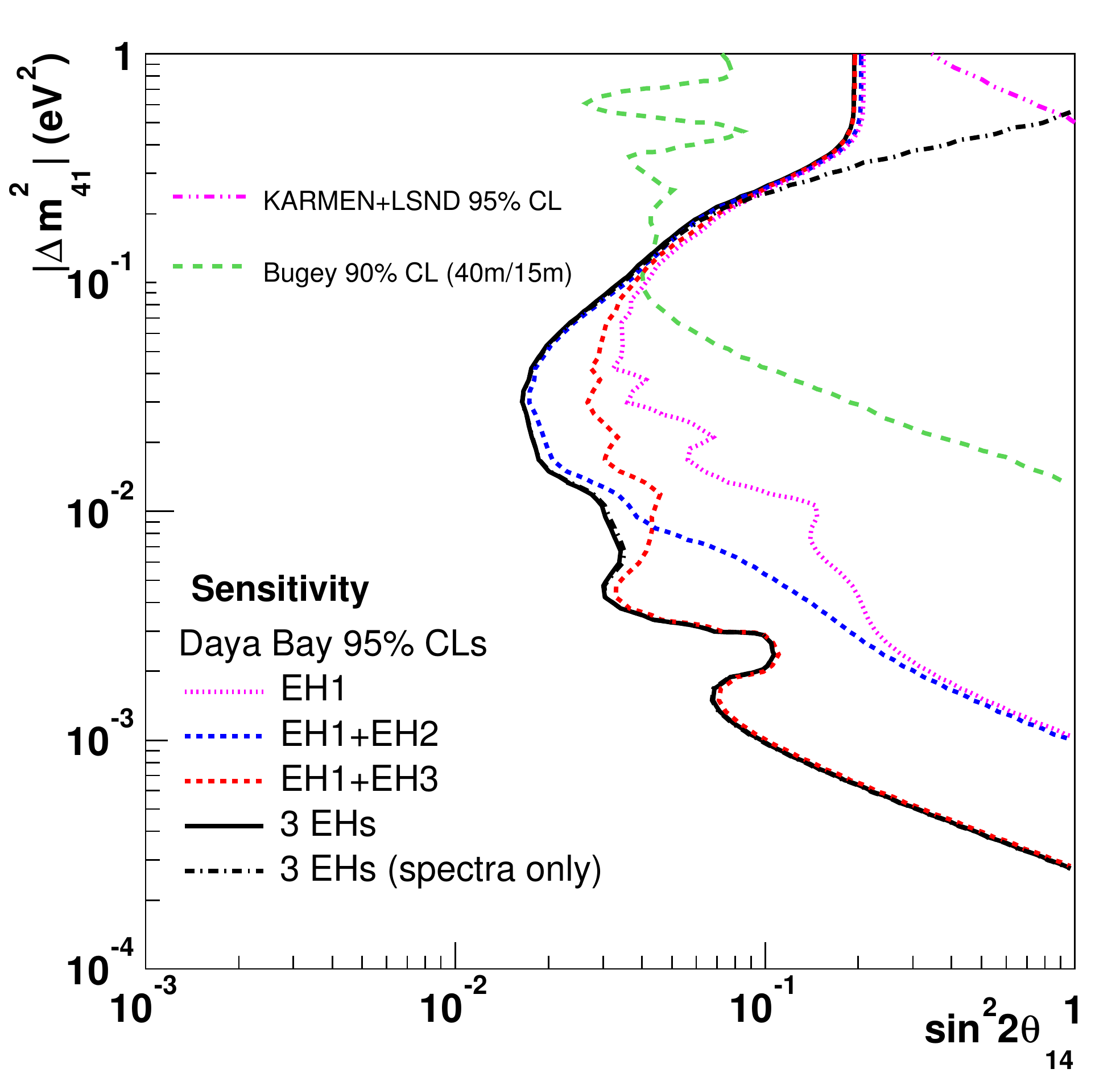}
\caption{
(color online) Comparison of the 95\% ${\rm CL_{s}}$ sensitivities (see text for details) for various combinations of the EH's data. The sensitivities were estimated from an Asimov Monte Carlo data set that was generated without statistical nor systematic variations. 
All the Daya Bay sensitivity curves were calculated assuming
5\% rate uncertainty in the reactor flux except the dot-dashed one, which 
corresponds to a comparison of spectra only.
Normal mass hierarchy was assumed for both $\Delta m_{31}^{2}$ and $\Delta m_{41}^{2}$. 
The dip structure at $|\Delta m^{2}_{41}| \approx 2.4 \times 10^{-3}~{\rm eV}^2$  was caused by the degeneracy between $\sin^{2}2\theta_{14}$ and $\sin^{2}2\theta_{13}$.
The green dashed line represents Bugey's~\cite{Declais:1994su} 90\% C.L. limit
on $\overline{\nu}_e$ disappearance and
the magenta double-dot-single-dashed line represents the combined KARMEN and LSND 95\% C.L. limit on $\nu_e$ disappearance 
from $\nu_e$-carbon cross section measurements~\cite{Conrad:2011ce}.}
\label{fig:sensitivity}
\end{figure}

\par 
 
The greatest sensitivity to $\sin^{2}2\theta_{14}$ in the   
$|\Delta m_{41}^{2}| < 0.3~{\rm eV}^2$ region came from 
the relative measurements between multiple EHs at different baselines.
Figure~\ref{fig:SpectralRatio} shows the ratios of the observed prompt energy spectra at EH2 (EH3) and the three-neutrino best fit prediction from the EH1 spectrum~\cite{An:2013zwz}.
The data are compared with the 3+1 neutrino oscillation with $\sin^{2}2\theta_{14} = 0.1$ and two representative $|\Delta m^2_{41}|$ values, illustrating that the sensitivity at $|\Delta m^2_{41}|=4 \times 10^{-2}$ $(4 \times 10^{-3})~{\rm eV}^2$ came primarily from the relative
spectral shape comparison between EH1 and EH2 (EH3). 
Sensitivities for various combinations of the data sets from different EHs were estimated with the method described later in this Letter, and are shown in Fig.~\ref{fig:sensitivity}.
The sensitivity in the $0.01~{\rm eV}^2 < |\Delta m_{41}^{2}| 
< 0.3~{\rm eV}^2$ region originated 
predominantly from the relative measurement between 
the two near halls, while the 
sensitivity in the $|\Delta m_{41}^{2}| < 0.01~{\rm eV}^2$ region arose primarily from the comparison between the near and far halls. 
The high-precision data at multiple baselines are essential for probing a wide range of values of $|\Delta m_{41}^{2}|$.
 
\par

The uncertainty of the reactor flux model's normalization had a marginal impact in the $|\Delta m_{41}^{2}| < 0.3~{\rm eV}^2$ region.
 For $|\Delta m_{41}^{2}| > 0.3~{\rm eV}^2$, spectral distortion features are smeared out and the relative measurement loses its discriminatory power. 
The sensitivity in this region can be regained by comparing  
the event rates of the Daya Bay near halls with the flux model 
prediction, which will be reported in a future publication.
In this Letter, we focus on the 
$|\Delta m_{41}^{2}| < 0.3~{\rm eV}^2$ region.

\par
Three independent analyses were conducted, each with a different treatment of the predicted reactor antineutrino flux and systematic errors. The first analysis used the predicted reactor antineutrino spectra to simultaneously fit the data from the three halls, in a fashion similar to what was described in the recent Daya Bay spectral analysis~\cite{An:2013zwz}. 
A binned log-likelihood method was adopted with nuisance parameters constrained with the detector response and the backgrounds, and with a covariance matrix encapsulating the reactor flux uncertainties  as given in the Huber~\cite{Huber:2011wv} and Mueller~\cite{Mueller:2011nm} flux models.
The rate uncertainty of the absolute reactor $\overline{\nu}_{e}$ flux was enlarged to 5\% based on Ref.~\cite{Hayes:2013wra}. The fit used $\sin^{2}2\theta_{12} = 0.857 \pm 0.024$, $\Delta m_{21}^{2} = (7.50 \pm 0.20)\times 10^{-5}~{\rm eV}^{2}$~\cite{Gando:2010aa} and $|\Delta m_{32}^{2}| = (2.41 \pm 0.10)\times 10^{-3}~{\rm eV^{2}}$ ~\footnote{This value was reported in Ref.~\cite{Adamson:2013whj}. An independent measurement was recently released in Ref.~\cite{T2K:2014disap}. Both values are consistent, and the results presented here are not very sensitive to this parameter.}.
The values of $\sin^{2}2\theta_{14}$, $\sin^{2}2\theta_{13}$ and $|\Delta m_{41}^{2}|$ were unconstrained. 
For the 3+1 neutrino model, a global minimum of $\chi_{4\nu}^{2} / {\rm NDF} = 158.8 / 153$ was obtained, 
while the minimum for the three-neutrino model was $\chi_{3\nu}^{2} / {\rm NDF} = 162.6 / 155$. We used the $\Delta \chi^{2} = \chi_{3\nu}^{2} - \chi_{4\nu}^{2}$ distribution obtained from three-neutrino Monte Carlo samples that incorporated both statistical and systematic variations to obtain a p-value~\footnote{The p-value is the probability of obtaining a test statistic result at least as extreme as the observed one.} of 0.74 for $\Delta \chi^{2} = 3.8$. The data were thus found to be consistent with the three-neutrino model, and there was no significant evidence for sterile neutrino mixing. 

\par

The second analysis performed a purely relative comparison between data at the near and far halls. The observed prompt energy spectra of the near
halls were extrapolated to the far hall and compared with observation. This process was done independently for each prompt energy bin, by first unfolding it into the corresponding true
antineutrino energy spectrum and then extrapolating to the far hall
based on the known baselines and the reactor power profiles.  
A covariance matrix, generated from a large Monte Carlo dataset incorporating both statistical and systematic 
variations, was used to account for all uncertainties. The resulting p-value was 0.87. More details about this approach 
can be found in Ref.~\cite{BerkeleyFitter}. 

\par  
The third analysis exploited both rate and spectral information in a way that is similar to the first method but using 
a covariance matrix. This matrix was calculated based on standard uncertainty propagation methods, without an extensive generation of Monte Carlo samples. The obtained p-value was 0.74.

\par
The various analyses have complementary strengths. Those that incorporated reactor antineutrino flux constraints had a slightly higher reach in sensitivity, particularly for higher values of $|\Delta m^2_{41}|$. The purely relative
analysis was more robust against uncertainties in the
predicted reactor antineutrino flux. 
The different treatments of systematic
uncertainties provided a thorough cross-check of the results, which were found to be consistent for all the analyses in the region where the relative spectral measurement dominated the sensitivity ($|\Delta m^2_{41}| < 0.3~{\rm eV}^2$). As evidenced by the reported p-values, no significant signature for sterile neutrino mixing was found by any of the methods. 

\par
Two methods were adopted to set the exclusion limits in the
$(|\Delta m^{2}_{41}|,\sin^{2}2\theta_{14})$ space.
The first one was a frequentist approach with a likelihood ratio as the ordering principle, as proposed by Feldman 
and Cousins~\cite{Feldman:1997qc}. For each point $\boldsymbol{\eta} \equiv (|\Delta m^{2}_{41}|,\sin^{2}2\theta_{14})$, the value $\Delta\chi^2_{c}(\boldsymbol{\eta})$ encompassing a fraction $\alpha$ of the events in the $\chi^2(\boldsymbol{\eta})- \chi^2(\boldsymbol{\eta}_{\rm best})$ distribution was determined, where $\boldsymbol{\eta}_{\rm best}$ was the best-fit point. This distribution was obtained by fitting a large number of simulated experiments that included statistical and systematic variations. To reduce the number of computations, the simulated experiments were generated with a fixed value of $\sin^22\theta_{13}=0.09$~\cite{An:2013zwz}, after it was verified that the dependency of $\Delta\chi^2_{c}(\boldsymbol{\eta})$ on this parameter was negligible. The point $\boldsymbol{\eta}$ was then declared to be inside the $\alpha$ C.L. acceptance region if $\Delta\chi^2_{\rm data}(\boldsymbol{\eta}) < \Delta\chi^2_{c}(\boldsymbol{\eta}).$

\begin{figure}[htb]
\includegraphics[width=\columnwidth]{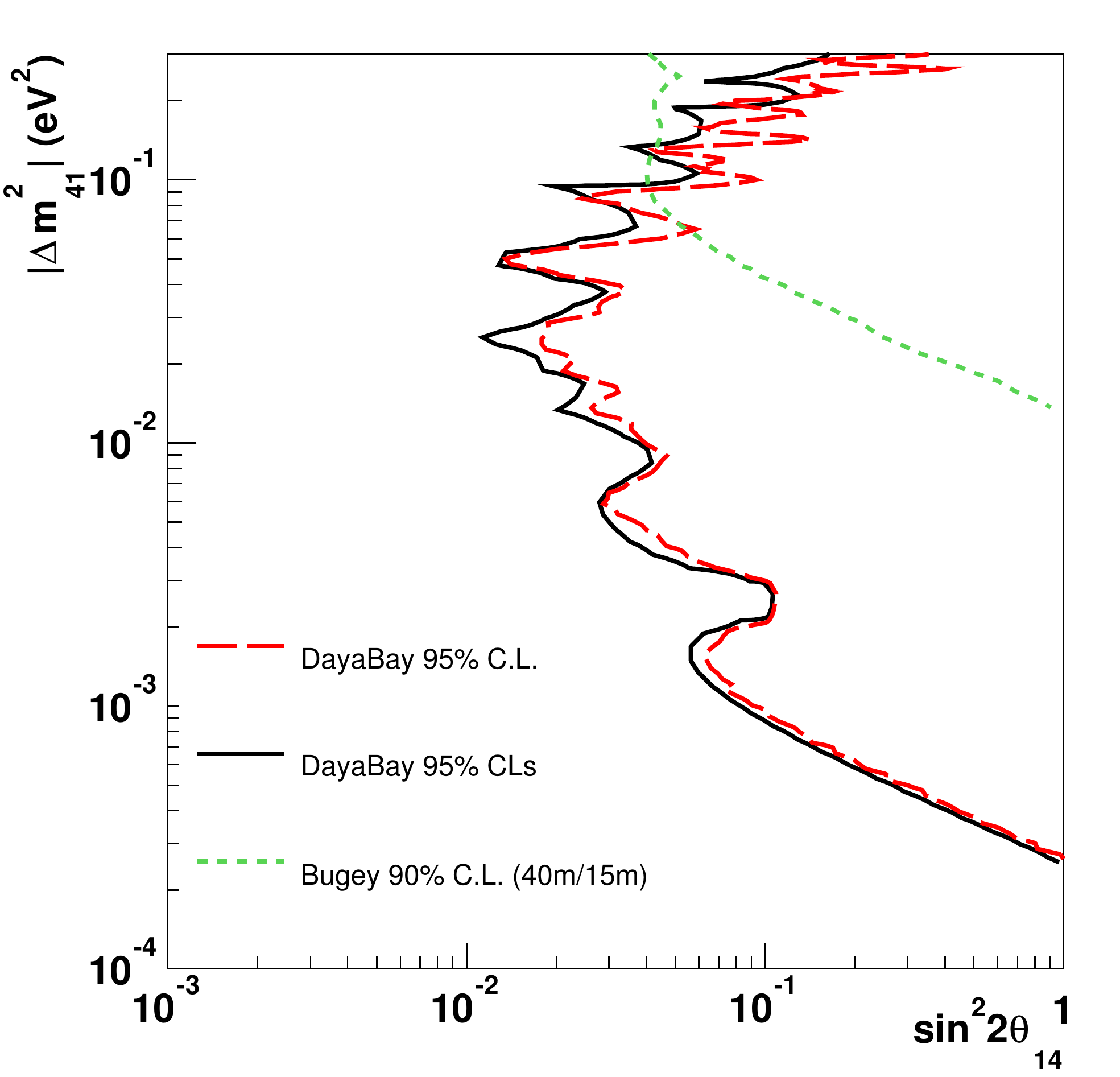}
\caption{
(color online) The exclusion contours for the neutrino oscillation parameters
  $\sin^{2}2\theta_{14}$ and $|{\Delta}m^{2}_{41}|$.
Normal mass hierarchy is assumed for both $\Delta m_{31}^{2}$ and $\Delta m_{41}^{2}$. 
The red long-dashed curve represents the 95\% C.L.  
exclusion contour with Feldman-Cousins method~\cite{Feldman:1997qc}.
The black solid curve represents the 95\% ${\rm CL_{s}}$ exclusion  
contour~\cite{Read:2002hq}. 
The parameter-space to 
the right side of the contours are excluded.
For comparison,
Bugey's~\cite{Declais:1994su} 90\% C.L. limit on $\overline{\nu}_e$ disappearance 
is also shown as the green dashed curve. 
 \label{fig:excludeOsc}}
\end{figure}

\par
The second method was the ${\rm CL_{s}}$ 
statistical method~\cite{Read:2002hq} 
described in detail in Ref.~\cite{CLsMethod}. 
A two-hypothesis test was performed
in the ($\sin^{2}2\theta_{14}$, $|\Delta m^{2}_{41}|$) phase space with
the null hypothesis $H_{0}$ (3-$\nu$ model) 
and the alternative hypothesis $H_{1}$ (3+1-$\nu$ model 
with fixed value of $\sin^{2}2\theta_{14}$ and $|\Delta m^{2}_{41}|$).
The value of $\theta_{13}$ was fixed with the best-fit value of the data for each hypothesis.
Since both hypotheses have fixed values of 
$\sin^{2}2\theta_{14}$ and $|\Delta m^{2}_{41}|$, 
their $\chi^{2}$ difference follows a Gaussian distribution. 
The mean and variance of these Gaussian distributions were 
calculated from Asimov datasets 
without statistical or systematic fluctuations, 
which avoided massive computing.  
The ${\rm CL_{s}}$ value is defined by:  
\begin{equation}
{\rm CL_{s}} = \frac{1-p_{1}}{1-p_{0}},
\end{equation}
\noindent where $p_{0}$ and $p_{1}$ 
are the p-values for the 3-$\nu$ and 3+1-$\nu$ 
hypotheses models respectively. 
The condition of ${\rm CL_{s}} \leq 0.05$ was required to set 
the 95\% ${\rm CL_{s}}$ exclusion regions.

\par
The 95\% confidence level contour from the Feldman-Cousins 
method and the 95\% ${\rm CL_{s}}$ method exclusion contour are shown 
in Fig.~\ref{fig:excludeOsc}. 
The two methods gave comparable results. The detailed structure is due to the finite statistics of the data. The impact of varying the bin size of the IBD prompt energy spectrum from 200 keV to 500 keV was negligible.
Moreover, the choice of mass ordering in both the three- and four-neutrino scenarios had a marginal impact on the results.
For comparison, Bugey's 90\% C.L. exclusion on $\overline{\nu}_e$ disappearance obtained from their ratio of the positron energy spectra measured at 40/15~m ~\cite{Declais:1994su} is also shown.
Our result presently provides the most stringent limits on sterile neutrino mixing at $|\Delta m^2_{41}| < 0.1~{\rm eV}^2$ using the electron antineutrino disappearance channel. This result is complementary to those from the $\pbar{\nu}_\mu \rightarrow \pbar{\nu}_e$ and $\pbar{\nu}_\mu \rightarrow \pbar{\nu}_\mu$ oscillation channels. 
While the $\pbar{\nu}_{e}$ appearance mode constrains the product of $|U_{\mu4}|^2$ and $|U_{e4}|^2$,  the $\pbar{\nu}_\mu$ and $\pbar{\nu}_{e}$ disappearance modes constrain $|U_{\mu4}|^2$ and $|U_{e4}|^2$, respectively.

\par

In summary, we report on a sterile neutrino search based on 
a minimal extension of the Standard Model, the 3 (active) + 1 (sterile) neutrino mixing model, 
in the Daya Bay Reactor Antineutrino Experiment
using the electron-antineutrino disappearance channel. 
The analysis used the relative event rate and the spectral comparison 
of three far and three near antineutrino detectors 
at different baselines from six nuclear reactors.
The data are in good 
agreement with the 3-neutrino model. 
The current precision is dominated by statistics.
With at least three more years of additional data, the
sensitivity to $\sin^22\theta_{14}$ is expected to improve by a factor
of two for most $\Delta m^2_{41}$ values. The current result already yields the world's most stringent limits on $\sin^{2}2\theta_{14}$ in the $|\Delta m_{41}|^{2} < 0.1$ eV$^{2}$ region. 

\par

Daya Bay is supported in part by the Ministry of Science
and Technology of China, the United States Department of
Energy, the Chinese Academy of Sciences, the National Natural Science Foundation of China,
the Guangdong provincial government, the Shenzhen municipal government, the
China Guangdong Nuclear Power Group,
Key Laboratory of Particle \& Radiation Imaging (Tsinghua University), Ministry of Education,
Key Laboratory of Particle Physics and Particle Irradiation (Shandong University), Ministry of Education,
Shanghai Laboratory for Particle Physics and Cosmology, the Research Grants
Council of the Hong Kong Special Administrative Region of
China, University Development Fund of The University of
Hong Kong, the MOE program for Research of Excellence at
National Taiwan University, National Chiao-Tung University,
and NSC fund support from Taiwan, the U.S. National Science Foundation, the Alfred P. Sloan Foundation, the Ministry
of Education, Youth and Sports of the Czech Republic, the Joint Institute of Nuclear Research in Dubna, Russia, the CNFC-RFBR joint research program,
National Commission of Scientific and Technological Research of Chile,
and Tsinghua University Initiative Scientific Research Program. We acknowledge Yellow River Engineering Consulting Co., Ltd. and China railway 15th Bureau Group Co.,
Ltd. for building the underground laboratory. We are grateful for the ongoing cooperation from the China General
Nuclear Power Group and China Light \& Power Company.

\bibliographystyle{apsrev4-1}

\bibliography{sterile}


\begin{figure*}[htb]
\includegraphics[width=\columnwidth]{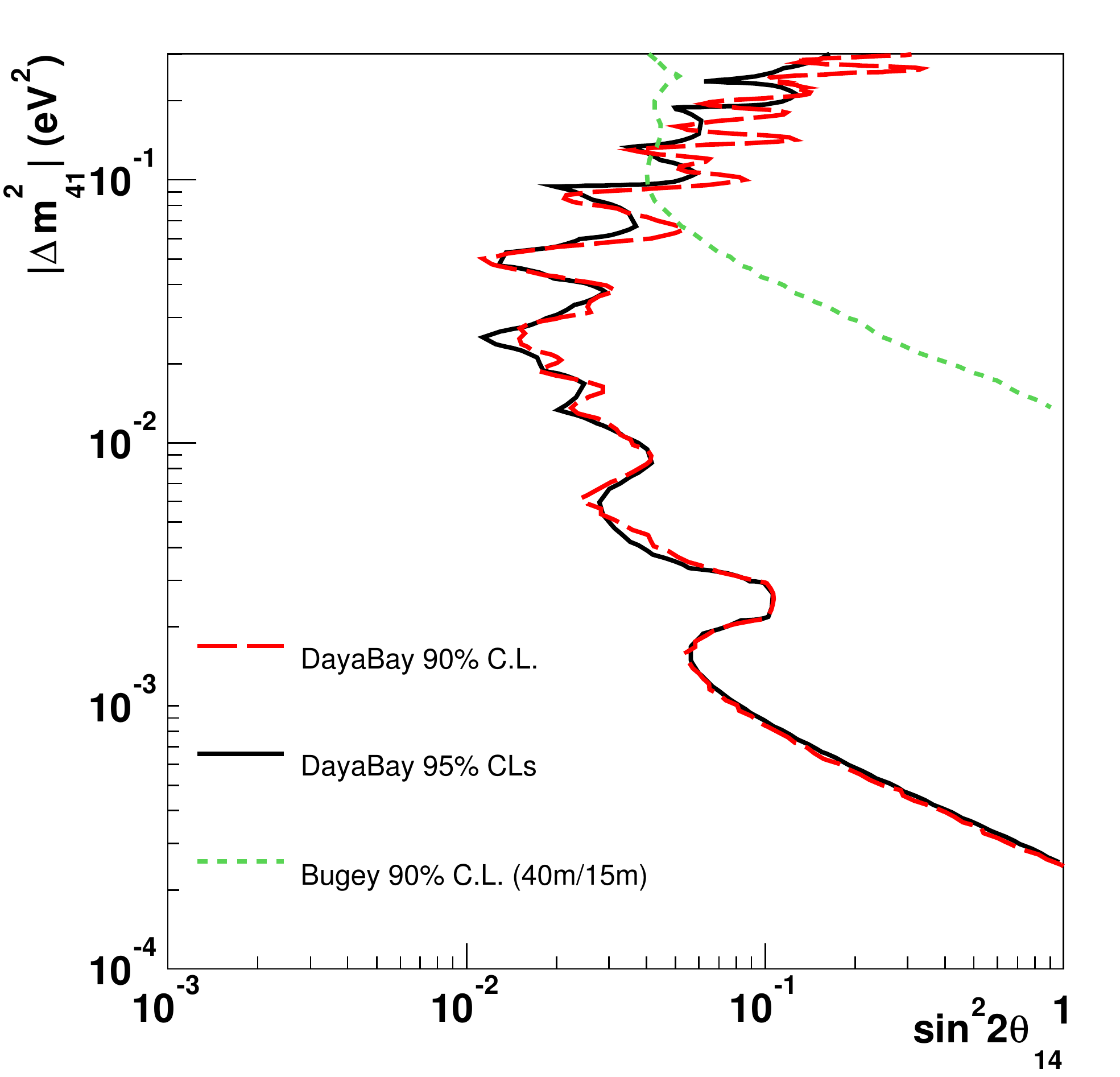}
\caption{
Same as Fig.~\ref{fig:excludeOsc} except that the 90\% C.L. exclusion contour is shown for the Feldman-Cousins method (supplemental material). 
 \label{fig:excludeOsc-90}}
\end{figure*}

\end{document}